\documentclass[doublecol]{epl2}

\newcommand{\be}{\begin{equation}}
\newcommand{\ee}{\end{equation}}
\newcommand{\bs}{\begin{subequations}}
\newcommand{\es}{\end{subequations}}

\title{Finite Size Dependent Dispersion Potentials Between Atoms and Ions Dissolved in Water}
\shorttitle{Finite Size Dependent Dispersion Potentials}

\author{P. Thiyam\inst{1}\and M. Bostr{\"o}m \inst{1,2,3} \and C. Persson \inst{1,2,4} \and D. F. Parsons \inst{5} \and I. Brevik \inst{3}  \and Bo E. Sernelius\inst{6}}

\institute{ \inst{1}Dept of Materials Science and Engineering, Royal Institute of Technology, SE-100 44 Stockholm, Sweden, EU\\
  \inst{2}Centre for Materials Science and Nanotechnology, University of Oslo, P. O. Box 1048 Blindern, NO-0316 Oslo, Norway\\
 \inst{3}Department of Energy and Process Engineering, Norwegian University of Science and Technology, NO-7491 Trondheim, Norway\\
 \inst{4}Department of Physics, University of Oslo, P. O. Box 1048 Blindern, NO-0316 Oslo, Norway\\
 \inst{5}Department of Applied Mathematics, Australian National University, Canberra, Australia\\
 \inst{6}Division of Theory and Modeling, Department of Physics,  Chemistry and Biology, Link\"{o}ping University, SE-581 83 Link\"{o}ping, Sweden, EU}

 \shortauthor{P. Thiyam \etal}
\pacs{34.20.Cf}{Interatomic potentials and forces}
\pacs{03.70.+k}{Theory of quantized fields}
\pacs{42.50.Lc}{Quantum fluctuations, quantum noise, and quantum jumps}

\abstract{
 A non-expanded theory is used for dispersion potentials between atoms and ions dissolved in a medium. The first order dispersion interaction between two atoms in an excited state must account for the fact that the two atoms are coupled via the electromagnetic field and must include effects from background media, retardation and finite size.  We show that finite size corrections when two particles are close change the dispersion interactions in water by several orders of magnitude. We consider as four illustrative examples helium atoms, krypton atoms, phosphate ions, and iodide ions.  We demonstrate that due to large cancellation effects retardation dominates the interaction for helium atom pairs in an isotropic excited state down to the very small atom-atom separations where finite size corrections are also important.}

\begin{document}
\maketitle

We aim in the present letter at calculating the Casimir-Polder interaction energy between atoms (helium and krypton) 
and ions (phosphate and iodine) dissolved in water and to study what effects the finite atomic size has on 
this interaction. We also give results for the resonance interaction energy between He atoms. The calculations presented are for very small interatomic distances (lower than 
ten {\AA}ngstr{\"o}ms) at which effects beyond dipole-dipole dispersion interactions are important.
The interaction energy between ground state helium and krypton atoms, in vacuum, are well known from standard ab initio calculations. They show a potential energy minimum ($\rho_{min}$=2.97{\AA}, $U_{min}$=0.001 eV for He$_2$, and $\rho_{min}$=4 {\AA}, $U_{min}$=0.016 eV for Kr$_2$, where $\rho$ is the atom-atom separation and $U$ is the interaction potential).  The minimum is the result of a repulsive short-range  electro-static part and an attractive long-range part, due electron correlation. We are here concerned with the attractive long range part  of the potential only. The formalism we use focuses on the dipole-dipole dispersion interaction. At very small separation higher order multipolar contributions contribute and when there is substantial wave-function overlap one has to resort to quantum-chemistry calculations since the electrons are no longer confined to their respective particle. In our formalism we find that the interaction stays finite when the particles come close to each other. We  emphasize that the reason  is that the particle radii have been assumed to be finite.  

There are recent developments in the literature related to our approach. Thus Kysylychyn {\it et al.} \cite{kysylychyn13} have studied  the interaction between a finite-size nanoparticle (could be an atom) and the surface of a solid, making use of the local-field method. Assuming that the nanoparticle has a finite nonlinear polarizability, an interaction potential is derived that is repulsive at short ranges and has an attractive long-range tail. Moreover, Przybytek {\it et al.} \cite{przybytek10, przybytek12} have done high precision work on the retarded Casimir-Polder potential between two  He atoms. When {$\rho > 100$\AA} retardation makes the diatomic potential switch from the London $\rho^{-6}$ decay to the Casimir-Polder $\rho^{-7}$ form. The retardation effect is found to increase the size of the helium dimer bound state by about 5\%. The nonretarded calculations yield an average value of $\rho$ for the helium bound state to be quite large, 47.1{\AA}, which agrees fairly  well with the measured  (retarded) value of 52{\AA}.  It may also be mentioned that an extensive treatment of quantum electrodynamical (QED) interactions and thermophysical properties of helium have  recently been given by Cencek {\it et al.} \cite{cencek12}. A recent paper by DiStasio {\it et al.}\,\cite{DiStasio } dealing with density functional (quantum mechanical) theory of the interaction potential between finite-sized quantum harmonic oscillators explicitly takes orbital overlap into account.  They find results  that are similar to ours in water. In both approaches finite atomic size renders the interaction finite at zero atom-atom separation. What is evident in our theory is the removal of the divergences at zero separation.  

In this letter we treat both the Casimir-Polder interaction and the resonance interaction. We begin by giving a brief introduction to resonance interaction. Ninham and co-workers\,\cite{Bostrom1}  demonstrated a decade ago that, due to some  drastic approximations, the underlying theory of resonance interactions in free space derived from perturbative QED is incorrect at large distances.\,\cite{Bostrom1,BostPRA2012,Hartman,Sherkunov} For small separations the QED result has been considered to be correct. However, the QED formalism supposes  a point dipolar description. One aim of this letter is to demonstrate how the taking into account of finite atomic size and background medium fundamentally alters the resonance interaction energy. In the formalism one atom is in its ground state and the other in an excited state. At resonance the excitation switches back and forth between the two atoms. The interaction can be separated into three branches, the $x$-, $y$-, and $z$-branch where the name of the branch denotes in what direction the oscillating dipoles are pointing. We let the $z$-direction be defined by the line joining the two atoms. When there is no preferred direction for the excitation all three branches are activated. We refer to this case as isotropically excited atom pairs.

From writing up the equations of motion for the excited system it is straightforward to derive the zero temperature Green's function for two identical (and isotropic) atoms dissolved in water\,\cite{McLachlan,Bostrom1}. The resonance condition\cite{Bostrom1} can be obtained from the following condition: $\tilde 1 - \alpha^* (\omega )^2 {\tilde T^2}(\rho |\omega ) = 0$, where $\tilde T$ is the susceptibility tensor and   $\alpha^* (\omega)$  the excess polarisability   of the atom dissolved in water. 
The excess polarisabilities ($\alpha^*(i \xi$)) and atomic radii ($a$) for helium and other noble gas atoms dissolved  in water were derived as in several papers by Parsons and Ninham\cite{ParsonsNinham2009,ParsonsNinham2010dynpol}. They were presented recently  by Bostr{\"o}m, Parsons, and co-workers.\,\cite{BosPar} Dynamic polarisabilities of noble gas atoms in vacuum were calculated
using \textsc{Molpro}\,\cite{MOLPRO2008} at a coupled cluster singles and
double (CCSD) level of theory. An aug-cc-pV6Z basis set
\cite{WoonDunning1994,PetersonDunning2002} was used for a selection of noble gas atoms (He, Ne and Ar while
aug-cc-pV5Z was used for Kr\,\cite{WilsonWoonPetersonDunning1999}). In this work we consider helium and krypton atoms and iodide and phosphate ions.
Polarisabilities,  $\alpha(i\xi)$, in vacuum were transformed to excess
polarisabilities, $\alpha^{*}(i\xi)$,  in water via the relation for a dielectric sphere embedded in a dielectric
medium\,\cite{BoroudjerdiKimEtAl2005,LandauLifshitz-ElectrodynContMedia-v8},  $  \alpha^{*} (i \xi) =  R^3  ( \varepsilon_a-\varepsilon_\mathrm{w}  )/( \varepsilon_a + 2\varepsilon_\mathrm{w}  )$. Here $\varepsilon_\mathrm{w}$ is the dielectric function of water and $R$ is the radius of the atom. $ \varepsilon_a$ is the effective dielectric
function of the atomic sphere, estimated  from the atomic polarisability in vacuum as $ \varepsilon_a (i \xi) = 1 + 4\pi\alpha (i \xi)/V$, where
$V$ is the volume of the atomic sphere.

In the case of two identical atoms the above resonance condition can be separated in one anti-symmetric and one symmetric part. Since the excited symmetric state has a much shorter life time than the anti-symmetric state the system can be trapped in an excited antisymmetric state.  The resonance interaction energy of this antisymmetric state can be evaluated by a simple  expression for  two dissolved atoms (excited in the $j$=$x$-, $y$-, or $z$-branch, where $z$ is in the direction of the line connecting the two atoms) in water,
\begin{equation}
{U_j}(\rho ) = \hbar \int\limits_{ - \infty }^\infty  {\frac{{d\xi }}{{2\pi }}} \ln \left[ {1 + \alpha^* (i\xi ){T_{jj}}(\rho |i\xi )} \right].
\label{Eq1}
\end{equation} 

The corresponding van der Waals (Casimir-Polder) interaction between (isotropic) atoms is given by the following expression:
\begin{equation}
{U^{CP}}(\rho ) = \frac{\hbar }{2}\sum\limits_{j = x,y,z} {\int\limits_{ - \infty }^\infty  {\frac{{d\xi }}{{2\pi }}} \ln \left[ {1 - \alpha^* {{(i\xi )}^2}{T_{jj}}{{(\rho |i\xi )}^2}} \right]},
\label{Eq2}
\end{equation}
where $\rho$ is the distance between the two atoms. Traditionally one assumes that the interaction is so weak that one may expand the logarithm in Eqs.\,(\ref{Eq1}) and (\ref{Eq2}) and keep the lowest order contribution only ($ \ln(1+x) \approx x$). In what follows we name this the expanded theory and when the full logarithmic expression is used the non-expanded theory.  


We now present the theoretical framework used to study how finite atomic size influences the Casimir-Polder interaction and the resonance interaction when two atoms are near each other. 
The polarisation cloud of real atoms has a finite spread and we consider as an interesting case (consistent with the modeling of the excess polarisability) a spatial distribution of the atom following a Gaussian function (for helium in water the radius $a$ is 0.60\AA).\,\cite{Mahanty2,Mahanty3}  We have, following the formalism developed by Mahanty and Ninham\,\cite{Mahanty2,Mahanty3}, derived the Green's function elements that account for retardation, background media, and finite size in an accurate way. They are
\begin{equation}
\begin{array}{*{20}{l}}
{{T_{xx}}\left( {\rho |i\xi } \right) = {T_{yy}}\left( {\rho |i\xi } \right)}\\
{ =  - \frac{{\exp \left[ {{{\left( {\frac{\xi }{c}} \right)}^2}{{\left( {\frac{a}{2}} \right)}^2}} \right]}}{{2\rho   }}\left\{ {\left[ {{{\left( {\frac{\xi }{c}} \right)}^2} + \left( {\frac{\xi }{c}} \right)\frac{1}{\rho } + {{\left( {\frac{1}{\rho }} \right)}^2}} \right]} \right.}\\
{\begin{array}{*{20}{l}}
{ \times \left[ {1 - {\rm{erf}}\left( {\frac{\xi }{c}\frac{a}{2} - \frac{\rho }{a}} \right)} \right]\exp \left( { - \frac{\xi }{c}\rho } \right)}\\
{ - \left[ {{{\left( {\frac{\xi }{c}} \right)}^2} - \left( {\frac{\xi }{c}} \right)\frac{1}{\rho } + {{\left( {\frac{1}{\rho }} \right)}^2}} \right]}
\end{array}}\\
{\begin{array}{*{20}{l}}
{ \times \left[ {1 - {\rm{erf}}\left( {\frac{\xi }{c}\frac{a}{2} + \frac{\rho }{a}} \right)} \right]\exp \left( {\frac{\xi }{c}\rho } \right)}\\
{\left. { - \frac{4}{{a\rho \sqrt \pi  }}\exp \left[ { - {{\left( {\frac{\xi }{c}} \right)}^2}{{\left( {\frac{a}{2}} \right)}^2} - {{\left( {\frac{\rho }{a}} \right)}^2}} \right]} \right\},}
\end{array}}
\end{array}
\label{Eq3}
\end{equation}
and
\begin{equation}
\begin{array}{*{20}{l}}
{{T_{zz}}\left( {\rho |i\xi } \right)}\\
{\begin{array}{*{20}{l}}
{ = \frac{{\exp \left[ {{{\left( {\frac{\xi }{c}} \right)}^2}{{\left( {\frac{a}{2}} \right)}^2}} \right]}}{  \rho }\left\{ {\left[ {\left( {\frac{\xi }{c}} \right)\frac{1}{\rho } + {{\left( {\frac{1}{\rho }} \right)}^2}} \right]} \right.}\\
{ \times \left[ {1 - {\rm{erf}}\left( {\frac{\xi }{c}\frac{a}{2} - \frac{\rho }{a}} \right)} \right]\exp \left( { - \frac{\xi }{c}\rho } \right)}
\end{array}}\\
{ + \left[ {\left( {\frac{\xi }{c}} \right)\frac{1}{\rho } - {{\left( {\frac{1}{\rho }} \right)}^2}} \right]\left[ {1 - {\rm{erf}}\left( {\frac{\xi }{c}\frac{a}{2} + \frac{\rho }{a}} \right)} \right]\exp \left( {\frac{\xi }{c}\rho } \right)}\\
{ - \left. {\frac{4}{{a\rho \sqrt \pi  }}\left( {1 + {{\left( {\frac{\rho }{a}} \right)}^2}} \right)\exp \left[ { - {{\left( {\frac{\xi }{c}} \right)}^2}{{\left( {\frac{a}{2}} \right)}^2} - {{\left( {\frac{\rho }{a}} \right)}^2}} \right]} \right\},}
\end{array}
\label{Eq4}
\end{equation}
where $c=c_0/\sqrt{\varepsilon(i\xi)}$ with $c_0$  the velocity of light in vacuum and $\varepsilon(i \xi)$ the dielectric function of water for imaginary frequencies.  The effect of the background medium is entirely contained in the velocity of light in the medium and the excess polarisability. In some papers a factor  $\varepsilon(i \xi)$ multiplies the excess polarisability and divides the Green's function elements. The results would of course be the same if we adapted that notation.
\begin{figure}
\includegraphics[width=8cm]{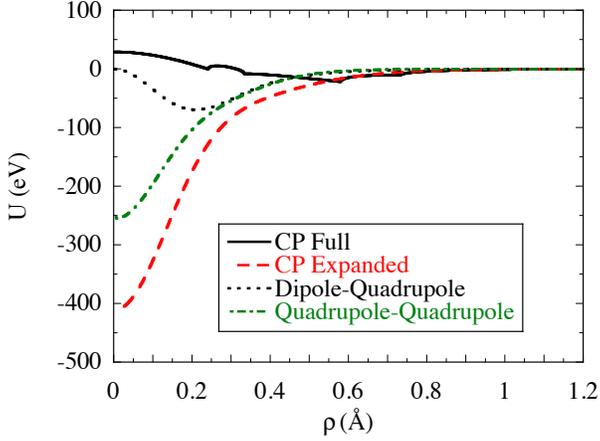}
\caption{(Color online) The van der Waals interaction between pairs of
finite size helium atoms in vacuum. Note that finite size effects keep the
interaction finite when the particles come close together. We also
show the van der Waals interaction between pairs of finite size helium
atoms when the theory is perturbatively 
expanded. We also show the curves for dipole-quadrupole and quadrupole-quadrupole contributions  in the expanded theory of dispersion energies.\,\cite{Richardson}}
\label{figu1}
\end{figure}
 \begin{figure}
\includegraphics[width=8cm]{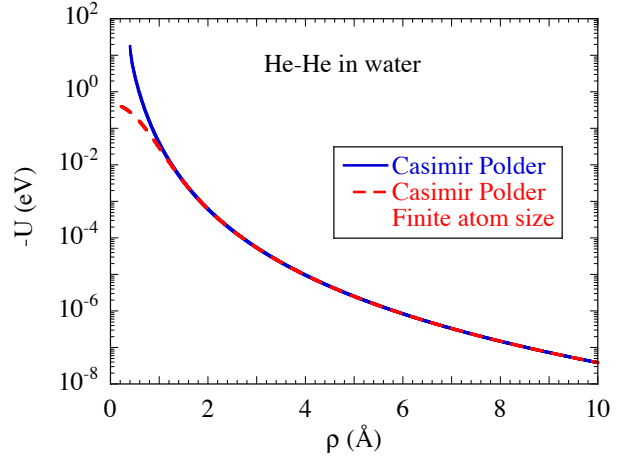}
\caption{(Color online) Casimir-Polder (ground-state) interaction between two helium atoms dissolved in water. Comparing theory with or without finite size dependence.}
\label{figu2}
\end{figure}

When retardation is neglected we find the following results:
\begin{equation}
\begin{array}{*{20}{l}}
{T_{xx}^{NR}(\rho |i\xi ) = T_{yy}^{NR}(\rho |i\xi )}\\
{ = \frac{{ - 1}}{{\sqrt \pi  {  \rho ^3}}}\left[ {\sqrt \pi  {\rm{erf}}\left( {\frac{\rho }{a}} \right) - 2\left( {\frac{\rho }{a}} \right){e^{ - {{\left( {\frac{\rho }{a}} \right)}^2}}}} \right],}
\end{array}
\label{Eq5}
\end{equation}
\begin{figure}
\includegraphics[width=8cm]{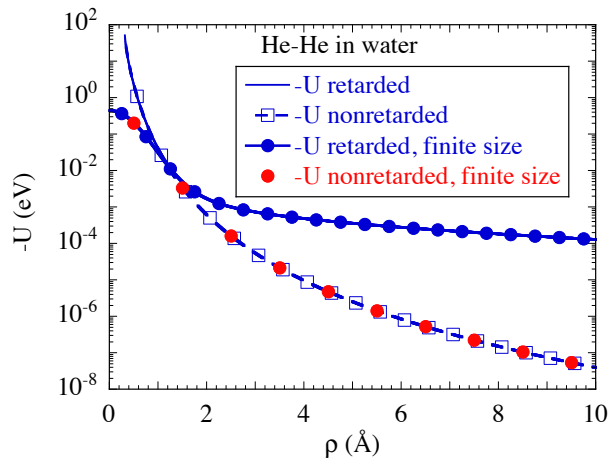}
\caption{(Color online)  Resonance interaction between isotropically excited diatomic helium dissolved in water. The effects of retardation and finite size ($a = 0.6$\,{\AA}) are considered. For distances smaller than the atomic radii retardation effects can be neglected.}
\label{figu3}
\end{figure}
\begin{figure}
\includegraphics[width=8cm]{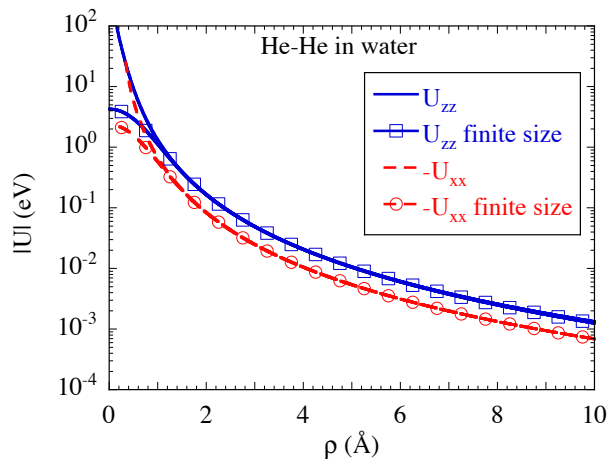}
\caption{(Color online) Resonance interaction between $x$- or $z$-mode in an excited state, with and without finite size effects included.  In this  plot  retarded and non-retarded results are indistinguishable. $U_{xx}$ is attractive, and its magnitude is shown here for comparison with $U_{zz}$. }
\label{figu4}
\end{figure}
\begin{equation}
\begin{array}{*{20}{l}}
{T_{zz}^{NR}(\rho |i\xi )}\\
{ = \frac{{ - 2}}{{\sqrt \pi  {  \rho ^3}}}\left[ { - \sqrt \pi  {\rm{erf}}\left( {\frac{\rho }{a}} \right) + 2\left( {\frac{\rho }{a}} \right)\left[ {1 + {{\left( {\frac{\rho }{a}} \right)}^2}} \right]{e^{ - {{\left( {\frac{\rho }{a}} \right)}^2}}}} \right],}
\end{array}
\label{Eq6}
\end{equation}
and
\begin{equation}
{\rm{Trace}}\left[ {{{\tilde T}^{NR}}\left( {\rho |i\xi } \right)} \right] = \frac{{ - 4}}{{\sqrt \pi  {a^3}}}{e^{ - {\rho ^2}/{a^2}}}.
\label{Eq7}
\end{equation}

We then find with either the $x$-branch or $z$-branch excited a $1/\rho^3$-dependence of the resonance interaction in the non-retarded limit. We observe that the resonance interaction at close contact depends on the radius $\propto a^{-3}$. This should be compared to the corresponding result for the ground state Casimir-Polder interaction that depends on the radius $\propto a^{-6}$. The analytical asymptotes are only valid for atomic systems where the logarithm can be series expanded.  

It is important to note that we are only considering the dispersive dipole interaction. We have not included contributions
due to  higher-order multipole contributions.\,\cite{Richardson,Buhmann,Duignan}  We show the resulting dipole-dipole, dipole-quadrupole, and quadrupole-quadrupole dispersion potentials\,\cite{Richardson}  in an expanded theory for two helium atoms in vacuum in  Fig.\,\ref{figu1} (note that the dipole-dipole interaction is a factor of 2 too small in Ref. \cite{Richardson} and the prefactor in their Eq.\,(34) should be $16/9\pi$ instead of $8/4\pi$). It is clear that dipole-dipole contributions give an important contribution at all separations. As comparison we also show the non-expanded result for  the dipole-dipole dispersion potential. Here we see that a non-expanded theory may account partially for the observed short-range repulsion between two atoms in vacuum. The fact that the atoms are in water will as we will see change this fact. 
In simulations that use a Lennard-Jones potential the interaction is made up of an attractive dispersion contribution and a repulsive contribution. We here calculate the attractive part in a correct way. A full quantum mechanical molecular simulation could be used to study how well our predictions represent the effect of water molecules on the interaction energy between two atoms at small interatomic distances. However, it should be observed that while a quantum chemistry calculation is an alternative way to model the interaction that too has its limitations. Quantum chemistry does not,  for example,  in an accurate way account for the coupling of atoms in an excited state. 

We show  in Fig.\,\ref{figu2}  the fully retarded finite temperature Casimir-Polder (ground-state) interaction between two helium atoms dissolved in water. Finite size corrections prevent the attraction from going to infinity when two atoms come very close together. This is consistent with the ground state Casimir-Polder interaction between finite size atoms in vacuum found by Mahanty and Ninham.\,\cite{Mahanty3}

In Fig.\,\ref{figu3}  we consider the effect of including retardation and finite size on the resonance interaction between isotropically excited diatomic helium dissolved in water. Finite size effects soften the attraction at very small separations but equally important is the effect of retardation.  One observes that for isotropically excited diatomic helium the non-retarded approximation which neglects finite size is not applicable at any separation. Both finite size dependence and retardation must be considered in order to have a theory that works for all separations.   For isotropically excited atom pairs the first expansion term in the logarithm cancels out and the leading non-retarded term is $\propto \rho^{-6}$. The dominating term is a retarded term $\propto \rho^{-4}$. 

We compare in Fig.\,\ref{figu4} the effect of finite size on the attractive $x$-mode excitation and repulsive $z$-mode excitation. It is seen that while finite size effects are conceptually important at very small atom-atom separations these effects play little role for separations beyond two atomic radii. Retardation effects are negligible for each of the contributing terms. However, there is a large cancellation effect (between the different $U_j (\rho)$ terms) when the contributions are added together and retardation is important for the resonance interaction between isotropically excited atom pairs as was seen in Fig.\,\ref{figu3}.  One important point in this letter is that while retardation reduces the $x$, $y$, and $z$ excited state resonance interactions ($1/\rho^3 \rightarrow 1/\rho^4$) the opposite is found for resonance interaction between isotropically excited atom pairs ($1/\rho^6 \rightarrow 1/\rho^4$).

Severe problems have been noted with the accepted theoretical expressions for the resonance interaction between identical atom pairs in an excited state when the atoms are far apart\,\cite{Bostrom1,BostPRA2012,Sherkunov}. Different research groups find very different results for the first order dispersion potential for atom pairs in an excited state at that separation limit. They do however find the same result when two atoms or molecules are close together.  
There are no debate concerning ground state interactions, nor for (anisotropic) resonance interactions in the non-retarded limit. However, our results for isotropically excited atom pairs propose that retardation (due to strong cancellations between $x$-, $y$- and $z$-components) could be important almost down to close contact. This may provide a case where experiments may be able to select the more appropriate way to calculate interactions between excited atom pairs. 
\begin{figure}
\includegraphics[width=8cm]{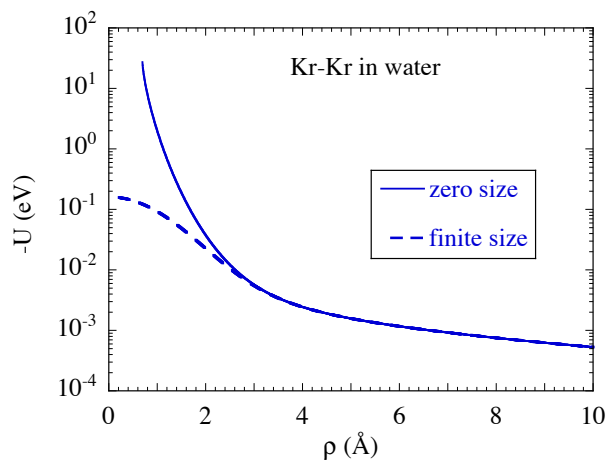}
\caption{(Color online) Retarded Casimir-Polder interaction between pairs of krypton atoms dissolved in water. Solid curve is for zero-size atoms and dashed for atoms of finite size. The finite radius of the krypton atoms is $a =1.4$\,{\AA}.}
\label{figu5}
\end{figure}
\begin{figure}
\includegraphics[width=8cm]{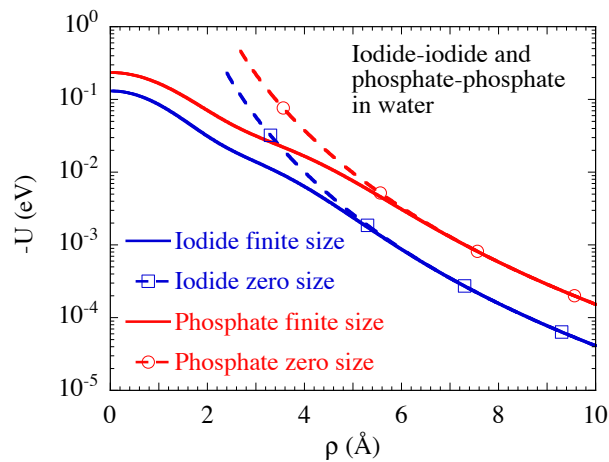}
\caption{(Color online) Retarded Casimir-Polder interaction between pairs of iodide ions (${{\rm{I}}^{\rm{ - }}}$) and pairs of phosphate ions (${\rm{PO}}_4^{3 - }$) dissolved in water.  The curves compare theory with and without finite size. The sizes for iodide and phosphate ions are $a =2.12$\,{\AA} and $a =2.40$\,{\AA}, respectively.}
\label{figu6}
\end{figure}

As we have demonstrated in this letter finite size effects that have been ignored  influence the resonance interaction between two atoms for separations of the order of a few atomic radii.  More substantial long range effects are expected for atoms, molecules and ions with larger Gaussian radii. We end by giving in Fig.\,\ref{figu5} and Fig.\,\ref{figu6} as further examples the resulting Casimir-Polder interaction for a pair of krypton atoms, a pair of phosphate ions, and a pair of iodide ions. Here, substantially longer ranged finite size effects are observed.



MB and CP acknowledge support from VR (Contract No. C0485101) and the Research Council of Norway (Project: 221469).  PT acknowledges support from the European Commission; this publication reflects the views only of the authors, and the Commission cannot be held responsible for any use which may be made of the information contained therein. MB also thanks the Department of Energy and Process Engineering (NTNU, Norway) for financial support.


\begin{thebibliography}{10}

\bibitem{kysylychyn13} D. Kysylychyn, V. Piatnytsia and V. Lozovski, Phys. Rev. E {\bf 88}, 052403 (2013).
\bibitem{przybytek10} M. Przybytek, W. Cencek, J. Komasa, G. Lach, B. Jeziorski and K. Szalewicz, Phys. Rev. Lett. {\bf 104}, 183003 (2010).
\bibitem{przybytek12} M. Przybytek, B. Jeziorski, W. Cencek, J. Komasa, J. B. Mehl and K. Szalewicz, Phys. Rev. Lett. {\bf 108 }, 183201 (2012).
\bibitem{cencek12} W. Cencek, M. Przybytek, J. Komasa, J. B. Mehl, B. Jeziorski and K. Szalewicz, J. Chem. Phys. {\bf 136}, 224303 (2012).
\bibitem{DiStasio} Jr R. A DiStasio, V. V Gobre and A. Tkatchenko, .J. Phys.: Condens. Matter, {\bf 26}  213202 (2014).


\bibitem{Bostrom1} Bostr\"{o}m M., Longdell J. J.,  Mitchell D. J., and Ninham B. W.,    {\it Eur. Phys. J. D} {\bf 22} (2003) 47. 
\bibitem{BostPRA2012}  Bostr\"{o}m M.,  Brevik I., Sernelius B. E.,  Dou M.,  Persson C., and Ninham B. W., {\it Phys. Rev. A} {\bf 86} (2012) 014701.  
\bibitem{Hartman} Hartman R. L.  and Leung P. T., {\it Phys. Rev. B} {\bf 64} (2001) 193308.
\bibitem{Sherkunov} Sherkunov Y.,  {\it Phys. Rev. A} {\bf  75} (2007)  012705.
 
\bibitem{McLachlan}  McLachlan A. D., {\it Molec. Phys.} {\bf 8} (1964) 409.
\bibitem{ParsonsNinham2009} Parsons D.F.  and  Ninham B. W., {\it J. Phys. Chem. A} {\bf 113} (2009) 1141.
\bibitem{ParsonsNinham2010dynpol} Parsons D.F.  and  Ninham B. W., {\it Langmuir} {\bf 26} (2010) 1816 .
\bibitem{BosPar} Bostr{\"o}m M., Ellingsen S. {\AA}., Brevik I.,  Parsons D. F.,  Sernelius Bo E., {\it Phys. Rev. A} {\bf 85} (2012) 064501.
\bibitem{MOLPRO2008} Werner  H.-J., et al., MOLPRO, version 2008.1, a package of ab initio programs, (2008). 
\bibitem{WoonDunning1994} Woon D. E.  and Dunning T. H.,  Jr., {\it J. Chem. Phys.} {\bf 100} (1994) 2975.
\bibitem{PetersonDunning2002} Peterson K. A.  and  Dunning T. H., Jr., {\it  J. Chem. Phys.} {\bf 117} (2002) 10548.
\bibitem{WilsonWoonPetersonDunning1999} Wilson A. K.,  Woon D. E., Peterson  K. A., and Dunning  T. H., Jr., {\it J. Chem. Phys.} {\bf 110} (1999) 7667.
\bibitem{BoroudjerdiKimEtAl2005} Boroudjerdi H., Kim Y.-W.,  Naji A.,  Netz R.R.,  Schlagberger X., and  Serr A., {\it Phys. Rep.} {\bf 416} (2005) 129.
\bibitem{LandauLifshitz-ElectrodynContMedia-v8}  Landau L. D. and  Lifshitz E. M., {\it Electrodynamics of Continuous Media}, (Pergamon Press, Oxford, 1960).

\bibitem{Mahanty2} Mahanty J.  and Ninham B. W., {\it Dispersion Forces} (Academic, London, 1976).
\bibitem{Mahanty3} Mahanty J.  and Ninham B. W., {\it J. Chem. Soc.  Faraday Trans. II} {\bf 71}  (1975) 119 (1975).
\bibitem{Richardson} D. D. Richardson, J. Phys. A: Math. Gen. {\bf 8}, 1828 (1975).

\bibitem{Buhmann} Buhmann S. Y., {\it Dispersion Forces. Macroscopic Quantum Electrodynamics and Ground-state Casimir, Casimir-Polder and Van der Waals Forces} (Springer-Verlag, Berlin and Heidelberg, 2013).
\bibitem{Duignan} Duignan  T.,  Parsons D. F., and  Ninham B. W., {\it J. Phys. Chem. B} {\bf 117}  (2013) 9412.

\end{thebibliography}
\end{document}